# Interlayer Coupling Effect in van der Waals Heterostructures of Transition Metal Dichalcogenides

Yuanyuan Wang, Fengping Li, Wei Wei,[*] Baibiao Huang, and Ying Dai[*]

*School of Physics, State Key Laboratory of Crystal Materials, Shandong University, Jinan 250100, China*

[*] Corresponding authors:

weiw@sdu.edu.cn

daiy60@sdu.edu.cn

Van der Waals (vdW) heterobilayers formed by two-dimensional (2D) transition metal dichalcogenides (TMDCs) created a promising platform for various electronic and optical properties. *ab initio* band results indicate that the band offset of type-II band alignment in TMDCs vdW heterobilayer could be tuned by introducing Janus WSSe monolayer, instead of an external electric field. On the basis of symmetry analysis, the allowed interlayer hopping channels of TMDCs vdW heterobilayer were determined, and a four-level $\boldsymbol{k}\cdot\boldsymbol{p}$ model was developed to obtain the interlayer hopping. Results indicate that the interlayer coupling strength could be tuned by interlayer electric polarization featured by various band offsets. Moreover, the difference in the formation mechanism of interlayer valley excitons in different TMDCs vdW heterobilayers with various interlayer hopping strength was also clarified.

**Keywords** *van der Waals heterostructures, transition metal dichalcogenides, interlayer*

*coupling effects, **k·p** model, interlayer exciton*

# 1 Introduction

In recent years, two-dimensional (2D) transition metal dichalcogenides (TMDCs) have attracted enormous attention as promising candidates for photonics, optoelectronics, and valleytronics [1, 2]. In regard to 2D TMDCs, the optical properties are characterized by tightly bound excitons, assigned to which the large exciton binding energy. In the light of inversion symmetry breaking, monolayer TMDCs present coupled spin–valley properties, which can be addressed by circularly polarized light excited excitons [3, 4]. It is inevitable in 2D monolayers that, however, sizeable electron–hole wave function overlap results in short (intralayer) exciton lifetimes and exchange interactions causes short valley polarization lifetimes due to the rapid valley mixing, being the major obstacles for the application of monolayer TMDCs in optoelectronics and valleytronics [5–8].

It is conclusive that the spin and valley degrees of freedom of charge carriers are associated with magnetic moments, while the layer degree of freedom in layered structures is relate to electric polarization. Therefore, combining two monolayer TMDCs into van der Waals (vdW) bilayer structures provides an unprecedented opportunity to realize the strong coupling between spin, valley as well as the layer pseudospin [9, 10]. It should be pointed out that the electronegativity difference between atomic layers may introduce an interlayer potential energy gradient in the TMDCs heterobilayers, which can cause appropriate energy band shift and type-II band alignment. In TMDCs bilayers with type-II band alignment, interlayer valley exciton forms, with electrons and holes finding their energy minima in opposite layers. In addition, large valley polarization will reduce the intarlayer hopping, therefore, suppresses the valley mixing. Thus, TMDCs heterobilayers present a promising platform for examining the emergent excitonic states and valleytronics [10–14].

In the absence of interlayer electric polarization, TMDCs vdW homobilayers usually show large interlayer interaction, *i.e.*, evident interlayer coupling effects, which results in strong interlayer hybridization. In practice, most naturally occurring TMDCs vdW bilayers are 2H stacking, where the upper layer is 180° rotated with respect to the lower layer [14]. As a result of the existed inversion symmetry, valley optical circular dichroism and valley Hall effect are vanishing in TMDCs vdW homobilayers. In this case, external electric field is usually introduced to break the inversion symmetry to realize valley polarization [15–17]. In contrast to vdW homobilayers of 2H phase, TMDCs heterobilayers show inherent inversion asymmetry, which plays a critical role in numerous physical phenomena such as the emergence of net valley polarization. In addition, with the formation of type-II band alignment, TMDCs heterobilayers possess ultrafast charge transfer and long-lived interlayer excitons (charge transfer excitons) under optical excitation, which is highly desirable for light–energy conversion applications [18, 19].

Consequently, most valleytronic studies involving interlayer excitons have focused on TMDCs vdW heterobilayers such as $WS_2/MoS_2$ and $MoSe_2/WSe_2$, and the tunability of the exciton lifetimes was confirmed by the external electric field [18–23]. It should be pointed out that the interlayer coupling effect is sensitive to the interlayer polarization, particularly, the large band offset will suppress the interlayer hopping. In TMDCs vdW heterobilayers, tunable interlayer polarization could be achieved without the introduce of external electric field. In this case, the low-energy exciton could be tuned from interlayer to intralayer by increasing the strength of interlayer polarization. It is therefore interesting to find TMDCs vdW heterobilayers with suitable band offset and interlayer hopping beyond the exiting systems to study the valley-selective excitonic states, and it is fundamentally significant to unravel their physical origin.

In this work, we systematically analyzed the symmetry features for interlayer coupling in three typical TMDCs vdW heterobilayers. In accordance to the atomic-layer sequence, these vdW heterobilayers are referred to as SeMoSe/SeWSe (S1), SeMoSe/SWSe (S2), and SeMoSe/SeWS (S3). In these prototypes, experimentally accessible TMDCs, monolayer Janus WSSe, was considered to include the dipole effects. In order to consider the influence of interlayer polarization on band alignment and interlayer coupling, energy dispersion relationship and interlayer hopping of these systems are obtained by $\boldsymbol{k\cdot p}$ model analysis in conjunction with the first-principles calculations. The formation process of interlayer exciton was clarified for SeMoSe/SeWSe and SeMoSe/SeWS heterobilayers. Our work proposes a reference for the further study on the interlayer valley excitons in TMDCs vdW heterobilayer systems.

## 2  Methods

The first-principles calculations were performed based on the density functional theory (DFT) using Vienna *ab initio* Simulation Package (VASP) [24]. All the calculations were carried out with Perdew–Burke–Ernzerhof (PBE) in the framework of generalized gradient approximation (GGA) [25] for the exchange–correlation functional, and the projector augmented wave (PAW) [26] was adopted for the electron–ion interactions. The spin–orbit coupling (SOC) was considered in the calculations. In order to mimic the isolated layers, a vacuum space was set to 20 Å. The cut-off kinetic energy for plane waves was set to 500 eV. Structures were fully relaxed until the force on each atom was less than 0.01 eV/Å$^{-1}$, and the convergence tolerance for energy was $10^{-5}$ eV. A Monkhorst–Pack *k*-point mesh of 13×13×1 was used to sample the Brillouin zone for geometry optimization and static electronic structure calculations [27]. In order to include the vdW interactions for the TMDCs heterobilayers, DFT–D2 corrections within the PBE functional were considered to obtain more accurate results [28].

## 3  Results and discussion

### 3.1 Interlayer hopping in TMDCs vdW bilayers

In consideration that certain interlayer hopping channels will vanish for configurations of special symmetry, the interlayer hopping preserved at $\pm K$ valleys for TMDCs bilayers with $\hat{C}_3$ rotation symmetry should be further discussed. The Bloch states at $\pm K$ valley is mainly contributed by transition metal $d_{z^2}$ and $\frac{1}{\sqrt{2}}(d_{x^2-y^2} \pm id_{xy})$ orbitals [29], therefore, the Bloch functions at band edges are

$$\psi_{\tau,\alpha}^L(r) = \frac{1}{\sqrt{N}} \sum_R e^{i\tau k \cdot R} d_m^L(r - R) \qquad (1)$$

$$\psi_{\tau',\beta}^U(r) = \frac{1}{\sqrt{N}} \sum_{R'} e^{i\tau' k \cdot R'} d_{m'}^U(r - R') \qquad (2)$$

where the summations are over all the metal sites $R$ in the lower layer (L) and $R'$ in the upper layer (U). N is the lattice site number of each layer, $\tau$ and $\tau'$ represent the valley indexes, $\alpha,\beta = \{c,v\}$ are the band indexes, and $m, m' = \{0, \pm 2\}$ are the magnetic quantum number of atomic orbits. The origin of coordinate is set on the site of a metal atom in the upper layer, thus $R' = l'a_1' + n'a_2'$ and $R = la_1 + na_2 - r_0$ with $a_{1,2}'$ and $a_{1,2}$ being the unit lattice vectors, and $r_0$ is the interlayer translation vector. The interlayer hopping between Bloch states at K valley is written as

$$t_{\beta\alpha} = \langle \psi_\beta^U | \hat{H} | \psi_\alpha^L \rangle = \frac{1}{N} \sum_{R,R'} e^{i k \cdot (R-R')} \int dr\, d_{m'}^U(r-R') \hat{H} d_m^L(r-R) \quad (3)$$

The interlayer hopping integral depends only on the relative position of two metals under a two-center approximation, with Fourier transformation the interlayer hopping matrix element is denoted as [30]

$$t_{\beta\alpha} = \frac{1}{N^2} \sum_{R,R'} e^{i k \cdot (R-R')} \sum_q T_m^{m'}(q) e^{-i q \cdot (R-R')} = e^{-i k \cdot r_0} \sum_\kappa T_m^{m'}(\kappa) e^{i \kappa \cdot r_0} \quad (4)$$

where $\kappa \equiv k + G$, and $G$ is the reciprocal lattice vector, $T_m^{m'}(\kappa)$ is the hopping integral. For 2H stacking of TMDCs at K point, $\kappa = k = K \equiv (\frac{4\pi}{3a}, 0)$, the global phase factor $e^{-i k \cdot r_0}$ can be dropped and $\hat{C}_3$ rotation simplifies the hopping

$$t_{\beta\alpha} = T_m^{m'} \left( e^{i K \cdot r_0} + e^{i \hat{C}_3 K \cdot r_0} e^{i \frac{2}{3}(m-m')\pi} + e^{i \hat{C}_3^2 K \cdot r_0} e^{-i \frac{2}{3}(m-m')\pi} \right) \quad (5)$$

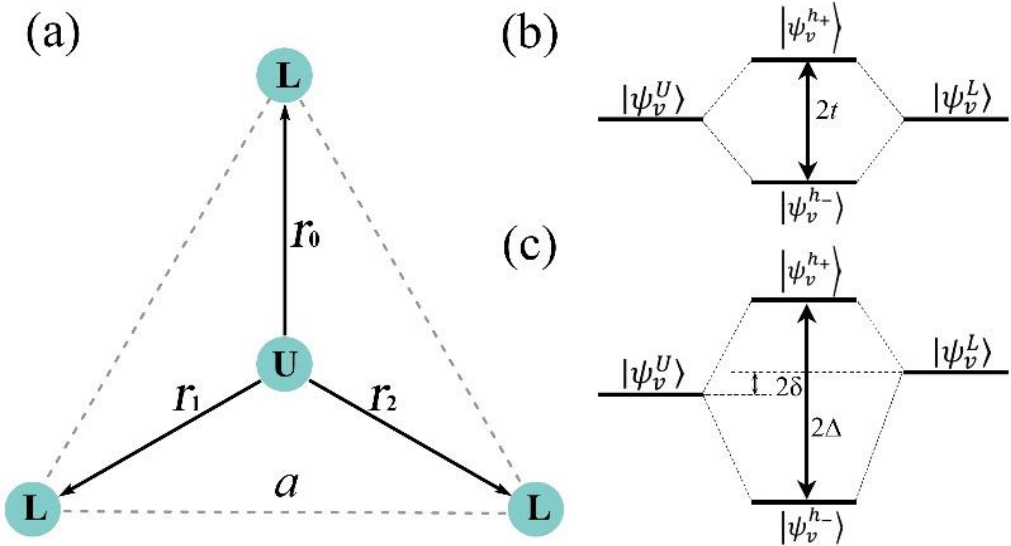

**Fig. 1.** (a) Nearest-neighbor interlayer hopping of metal atoms in 2H stacked TMDCs vdW heterobilayers, atoms located in upper (lower) layer are indicated by $U$ ($L$), $r_1$

and $r_2$ represent $\hat{C}_3 r_0$ and $\hat{C}_3^2 r_0$, respectively. Schematic showing the hybridization of electronic states between upper ($\langle\psi_v^U|$) and lower ($\langle\psi_v^L|$) layer in (b) homobilayer and (c) heterobilayer.

As shown in **Fig. 1a**, the nearest-neighbor interlayer hopping between transition atoms is considered, where only the interlayer hopping between valences band $t = t_{vv} \equiv 3T_2^{-2}$ is reserved. Such interlayer hopping in valance band causes the hybridization of electronic states from upper and lower layers of the vdW systems, see **Figs. 1b** and **1c**.

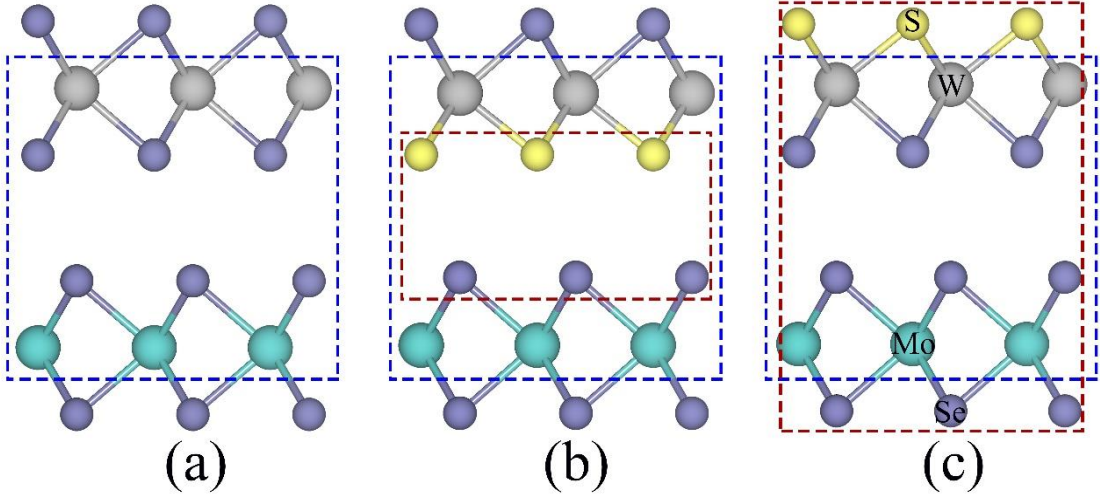

**Fig. 2** TMDCs vdW heterobilayers (a) SeMoSe/SeWSe (S1), (b) SeMoSe/SWSe (S2), and (c) SeMoSe/SeWS (S3). The blue (red) box denotes the permanent interlayer electric polarization $P_{\text{W-Mo}}$ ($P_{\text{S-Se}}$).

### 3.2 Electronic structures of TMDCs vdW heterobilayers

Once the interlayer hopping channel of 2H stacked TMDCs vdW heterobilayer is determined, the energy dispersion relationship of 2H stacked TMDCs vdW heterobilayers will be discussed. In order to obtain type-II band alignment and confirm whether the band offset will influence the interlayer hopping, three TMDCs vdW heterobilayers of 2H phase, *i.e.*, SeMoSe/SeWSe (S1), SeMoSe/SWSe (S2) and SeMoSe/SeWS (S3), were considered in this work. These TMDCs vdW heterobilayers show lattice mismatch less than 2.5%, which ensures the structure stability and easy experimental accessibility. As shown in **Fig. 2**, different interlayer potential differences lead to different interlayer electric polarization and type-II band alignment with different band offsets.

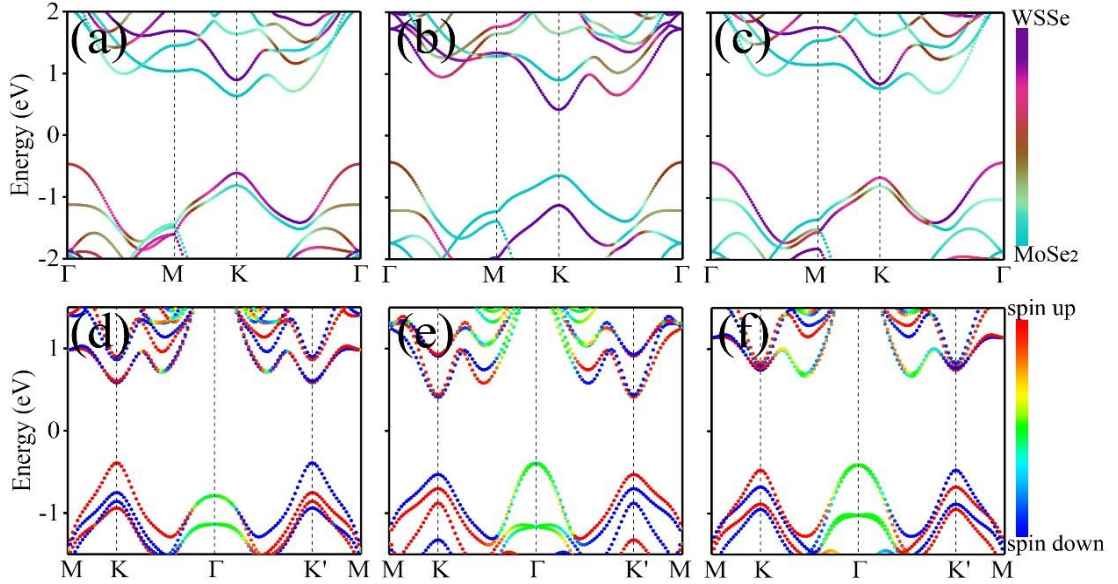

**Fig. 3** Layer-resolved band structures of (a) S1, (b) S2 and (c) S3, with the color-bar indicating the projection of the wave function onto either layer where spin-obit coupling (SOC) is not included. Spin-Resolved band structure of (d) S1, (e) S2 and (f) S3, with the color-bar denote the projection of spin direction. The Fermi level is set to zero.

In **Figs. 3a-c**, band structures without SOC are shown for of S1, S2 and S3 heterobilayers, with the color indicating the projection of wave function onto different layers. As expected, *ab initio* results confirm the type-II band alignment for all the heterobilayers considered. In particular, the valence band maximum (VBM) appears at the Γ point for all heterobilayers, while for S1 and S2 heterobilayers the conduction minimum (CBM) is located at K point and for S3 heterobilayer at a *k* point between K and Γ. In these heterobilayers, the SOC contribution can be simplified from $L \cdot S$ which can be confirmed by the projection of spin operator $\hat{s}_z$ at K/K' point. Thus, the spin-up and spin-down components are completely decoupled and $\hat{s}_z$ remains a good quantum number. In accordance to the spin-resolved band structures, SOC induces vertical spin splitting in VBM and the VBM shifts to K (K') point for S1 heterobilayer, as shown in **Fig. 3d**. Since SOC mainly contributes to K (K') point, thus the interlayer hybridization at K (K') point is stronger than that at Γ point. When SOC is considered, however, the reduced interlayer hybridization of S2 and S3 heterobilayers cannot induce large enough interband repulsion at K (K') point, thus the VBM remains at Γ point. It can be found that spin-valley-layer coupling occurs for three heterobilayers. In addition, electronic wave function projection shows that the VBM at K point is not 100% contributed by WSe$_2$ (WSSe) layer, but contributions of 8.8% (26.5%) are from MoSe$_2$ layer for S1 (S3), definitely indicating the interlayer coupling. However, for S2 heterobilayer, local VBM at K point is almost entirely contributed by MoSe$_2$ layer, due to the larger interlayer electric polarization $P_{\text{S-Se}}$ introduced by S-Se interface than the $P_{\text{W-Mo}}$ formed between atomic W and Mo layers, with $P_{\text{S-Se}}$ and $P_{\text{W-Mo}}$ exhibiting opposite direction.

### 3.3 Interlayer hopping integral from *k·p* model

In TMDCs vdW heterobilayers, the first valence and conduction bands near K point are dominantly contributed by $d_{z^2}$, $d_{xy}$, $d_{x^2-y^2}$ orbitals of metal atoms. Therefore, a four-level model near K point can be developed for the heterobilayers by adding the interlayer hopping to the *k·p* model for monolayers [31]

$$H(\boldsymbol{k}) = \begin{pmatrix} \epsilon_{u1} & at_u(\tau_z k_x + ik_y) & 0 & 0 \\ at_u(\tau_z k_x - ik_y) & \epsilon_{u2} & 0 & t \\ 0 & 0 & \epsilon_{l1} & at_l(\tau_z k_x - ik_y) \\ 0 & t & at_l(\tau_z k_x + ik_y) & \epsilon_{l2} \end{pmatrix} \quad (6)$$

The basis is $\{|d_{z^2}^u\rangle, 1/\sqrt{2}\left(|d_{x^2-y^2}^u\rangle - i\tau_z|d_{xy}^u\rangle\right), |d_{z^2}^l\rangle, 1/\sqrt{2}\left(|d_{x^2-y^2}^l\rangle + i\tau_z|d_{xy}^l\rangle\right)\}$, where the superscripts *u* and *l* denote the upper and lower layer, respectively. The wave vector *k* is measured around K (K') point, $\epsilon_{u(l)1} - \epsilon_{u(l)2}$ is the energy gap for upper

(lower) layer, $t$ is the interlayer hopping for holes, $t_u$ and $t_l$ are the nearest-neighbor intralayer hopping, $\tau_z$ is the valley index, $a$ is the lattice constant as shown in **Fig. 1a**. These parameters can be obtained by fitting the *ab initio* band structures. In accordance to our results, the interlayer hopping amplitude *t* corresponds to 46, 5 and 60 meV for S1, S2 and S3 heterobilayer, respectively. In comparison with previous works, in which $t_{vv}$ were calculated to be 45, 53 and 67 meV for MoSe$_2$/WSe$_2$, MoSe$_2$/MoSe$_2$, WSe$_2$/WSe$_2$ bilayers, respectively [21,31,32], our results are reasonable. It is an indication that interlayer electric polarization can substantially change the interlayer hopping, thus affecting the interlayer coupling effect.

In addition, the interlayer hybridization of electronic states can be simply described by a two-level Hamiltonian when the interband coupling is eliminated

$$H_v(\mathbf{k}) = \begin{pmatrix} \epsilon_{u2} & t \\ t & \epsilon_{l2} \end{pmatrix} \qquad (7)$$

The two layer-hybridized eigenvectors are $|\varphi_{h_+}\rangle = \mu|\varphi_u\rangle + \nu|\varphi_l\rangle$ and $|\varphi_{h_-}\rangle = \mu'|\varphi_u\rangle + \nu'|\varphi_l\rangle$, where $\mu/\nu = t/(\delta - \Delta)$, $\mu'/\nu' = t/(\delta + \Delta)$, $2\delta = \epsilon_{u2} - \epsilon_{l2}$ and $\Delta = \sqrt{\delta^2 + t^2}$ is the energy separation between the doublet, as depicted in **Figs. 1b** and **1c**. The degree of interlayer hybridization can be further defined as $P_H = |\mu^2/(\nu^2 + \mu^2)|$. From the *ab initio* calculation results, we determine the $P_H$ to be 91%, 1% and 73.5% for S1, S2 and S3 heterobilayer, respectively. It therefore further confirms that the interlayer electric polarization can significantly affect the strength of interlayer coupling. In other words, the strength of interlayer hybridization can be tuned by introducing a component layer of mirror asymmetry instead of a vertical gate field, that is, replacing TMDCs layer with mirror symmetry by Janus TMDCs.

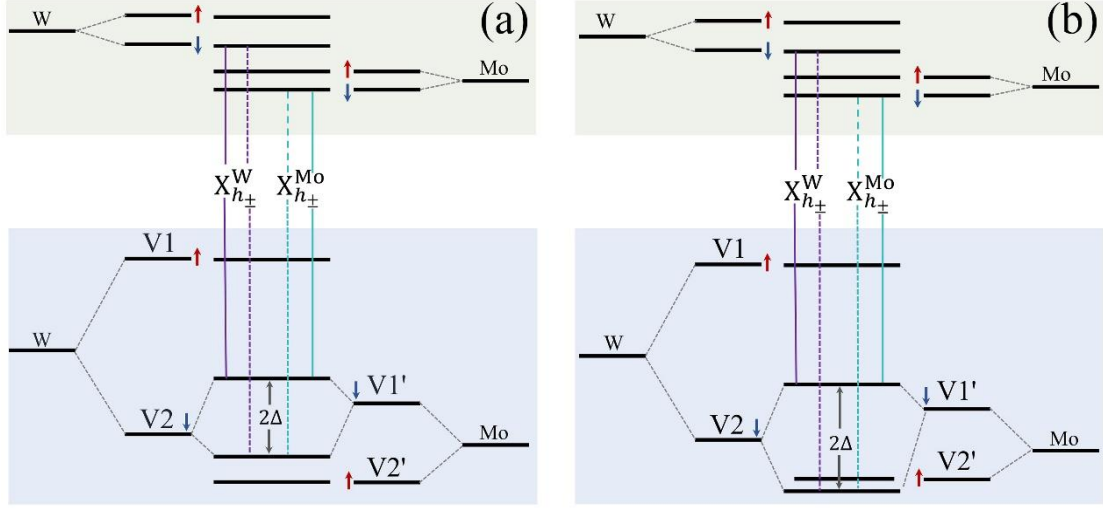

**Fig. 4** Schematic showing the optical transitions of 2H stacked (a) S1 heterobilayers and (b) S3 heterobilayers at the K valley. $X_B^W$ and $X_A^{Mo}$ transitions are split into two hybridized transitions $X_{h_\pm}^W$ (purple) and $X_{h_\pm}^{Mo}$ (green), respectively.

**3.4 Interlayer valley excitons in TMDCs vdW heterobilayers**

As discussed above, S1 and S3 heterobilayers exhibit type-II band alignment and larger interlayer hybridization than the S2 counterpart. As SOC is taken into account, spin-valley-layer coupling is reserved at the K point in S1 and S3 systems (see **Figs. 3d-f**). As shown in **Figs. 4a** and **4b**, the split-off valence bands are labeled as V1 and V2 for $WSe_2$ (WSSe) layer and V1' and V2' for $MoSe_2$ layer in S1 (S3) heterobilayer. Owing to the type-II band alignment and large SOC splitting for $WSe_2$ and WSSe, V2 band is located close to the V1' and V2' bands of $MoSe_2$. As discussed above, the symmetry of 2H stacking acquires zero interlayer hybridization ($t_{cc} = 0$) between conduction bands. Therefore, the interlayer coupling appears only in the valence band when taking spin into account, see **Figs. 4a** and **4b**.

On account of the spin conservation for interlayer hopping, V2 band of $WSe_2$ couples only to the V1' band of $MoSe_2$ in the case of S1 heterobilayer with middle strength of interlayer hopping. As a consequence, such a coupling hybridizes V2 and V1' bands into the doublet $|\psi_v^{h+}\rangle$ and $|\psi_v^{h-}\rangle$, as depicted in **Fig. 1c**. In regard to interband optical transition, the hybridized valance states ($|\psi_v^{h+}\rangle$ and $|\psi_v^{h-}\rangle$) can be excited into the conduction bands localized on WSe$_2$ and MoSe$_2$ layers, splitting the $X_B^W$ ($X_A^{Mo}$) exciton into the doublet, as shown in **Fig. 3d** and **Fig. 4a**, the $X_{h_+}^W$ and $X_{h_-}^W$ ($X_{h_+}^{Mo}$ and $X_{h_-}^{Mo}$). In comparison to S1 heterobilayer, S3 heterobilayer shows stronger interlayer hopping, thus, as can be found in **Fig. 3f** and **Fig. 4b**, conspicuous hybridization leads to the $|\psi_v^{h-}\rangle$ state lower in energy than V2' state. Thus, the four species of exciton consists of electron confined in an individual layer and a layer-hybridized hole, featuring a large electric dipole as well as a large optical dipole. In other words, these excitons combine the advantage of intralayer excitons of strong light coupling and interlayer excitons with strong dipole–dipole interactions and long lifetimes. As a consequence of the spatial indirect nature, the binding energy of interlayer excitons turns out to be small.

## 4  Summary

In summary, band structures of TMDCs vdW heterobilayers demonstrate that Janus WSSe introduced particular interlayer electric polarization can effectively tune the band offset. On the basis of the symmetric requirement of 2H stacked TMDCs vdW bilayers, allowed interlayer hopping at the neighborhood of K (K') point was obtain by fitted *ab initio* band results with the four-level ***k·p*** model. Results indicate that interlayer electric polarization will evidently affect the interlayer hopping strength, interlayer coupling strength can be tuned by changing the band offset without introducing external electric field. As a result of different interlayer coupling effect, interlayer exciton formation mechanism differs in heterobilayers with different interlayer hybridization.

**Acknowledgements**

This work is supported by the National Natural Science Foundation of China (No. 51872170), Young Scholars Program of Shandong University (YSPSDU), Shandong Provincial Key Research and Development Program (Major Scientific and Technological Innovation Project) (No. 2019JZZY010302), Natural Science Foundation of Shandong Province (No. ZR2019MEM013), and Taishan Scholar Program of Shandong Province.